\documentclass[aps,pra,amsmath,twocolumn,amssymb,superscriptaddress]{revtex4-1}

\usepackage{amssymb,amsfonts,amsmath}

\usepackage{color}
\usepackage[apple mac]{inputenc}
\usepackage[english]{babel}
\usepackage{times}
\usepackage{latexsym}
\usepackage{fancyhdr}
\usepackage{verbatim}
\usepackage{graphicx}
\usepackage{epsf}
\usepackage{subfigure}
\usepackage{bm}
\usepackage{epstopdf}\definecolor{Nathanblue}{rgb}{0.,0.24,0.51}

\definecolor{orange}{rgb}{0.96,0.24,0.00}

\def\be{\begin{equation}}
\def\ee{\end{equation}}

\begin{document}

\title{Quantized Rabi Oscillations and Circular Dichroism in Quantum Hall Systems
}

\author{D.~T. Tran}
\email[]{ducttran@ulb.ac.be}
\affiliation{Center for Nonlinear Phenomena and Complex Systems,
Universit\'e Libre de Bruxelles, CP 231, Campus Plaine, B-1050 Brussels, Belgium}
\author{N.~R. Cooper}
\affiliation{T.C.M. Group, Cavendish Laboratory, University of Cambridge, JJ Thomson Avenue, Cambridge, CB3 0HE, U.K.}
\author{N. Goldman}
\email[]{ngoldman@ulb.ac.be}
\affiliation{Center for Nonlinear Phenomena and Complex Systems,
Universit\'e Libre de Bruxelles, CP 231, Campus Plaine, B-1050 Brussels, Belgium}

\begin{abstract}

The dissipative response of a quantum system upon periodic driving can be exploited as a probe of its topological properties. Here, we explore the implications of such phenomena in two-dimensional gases subjected to a uniform magnetic field. It is shown that a filled Landau level exhibits a quantized circular dichroism, which can be traced back to its underlying non-trivial topology. Based on selection rules, we find that this quantized effect can be suitably described in terms of Rabi oscillations, whose frequencies satisfy simple quantization laws. We discuss how quantized dissipative responses can be probed locally, both in the bulk and at the boundaries of the system. This work suggests alternative forms of topological probes based on circular dichroism.

\end{abstract}

\date{\today}

\maketitle

\paragraph*{Introduction}

In the last decade, gases of ultracold atoms and photonic devices have revealed novel manifestations of topological states of matter, through the development of detection protocols that are complementary to those found in the solid-state context~\cite{Goldman_topology,Cooper_topology,Ozawa_review}. Examples of such experimental advances include the mapping of the Berry curvature~\cite{Duca_2015,Jotzu,Flaschner_2016,Wimmer:2017} and the extraction of  topological invariants (e.g.~Chern numbers)~\cite{Aidelsburger,Wu_2016,Lohse_2016,Nakajima,Sugawa_2016,Cardano_2017,Lohse_2018,Schine_2018} associated with Bloch bands, and the direct visualization of unidirectional edge modes~\cite{Wang_2009,Hafezi_2013,Rechtsman_2013,Stuhl_2015,Mancini_2015,Mukherjee_2017,An_2017}. 

Besides standard probes offered by the cold-atom toolbox~\cite{Bloch_review}, such as in-situ imaging of the particle density or the reconstruction of the momentum distribution through time-of-flight measurements, these systems also offer the possibility of monitoring energy absorption upon periodic driving~\cite{Eckardt_review}. Such ``heating" processes can be finely evaluated by measuring the dynamical repopulation of Bloch bands~\cite{Aidelsburger}, or through direct momentum-resolved multi-band spectroscopy~\cite{Flaschner_2018}.  These advances inspired the recent work~\cite{tran2017}, which showed how the topological properties of a quantum gas could be probed by analyzing such absorption processes upon applying a circular drive~\cite{Jotzu,Hamburg}; see also Refs.~\cite{Souza,Wang_circular_TI,Goldman_edge,Juan,Schuler_2017,YingLiu_2017} on connections between circular dichroism and topological properties. Specifically, a quantized dissipative effect was identified in Ref.~\cite{tran2017}, by investigating the depletion rate $\Gamma_{\pm} (\omega)$ of a 2D Chern insulator in response to a circular time-modulation,
\begin{align}
 \hat{V}_\pm(t)=2 E \left[\cos{(\omega t)}\hat{x}\pm\sin{(\omega t)}\hat{y}\right],
 \label{eq:circulardrive}
 \end{align}
 where $\hat{x}$ and $\hat{y}$ denote the position operators, $\omega$ is the drive frequency, and $\pm$ refer to the polarisation; such a drive can be realized by shaking a 2D optical lattice in a circular manner~\cite{Jotzu,Hamburg}. Using those notations, the quantized dissipative effect can be expressed as follows~\cite{tran2017}:
\be
\Delta \Gamma^{\text{int}}/A_{\text{syst}} = \eta_0 E^2,  \qquad   \eta_0=(1/\hbar^2) \,  \nu , \label{main_result}
\ee
where $\nu$ denotes the Chern number of the populated Bloch band,  $A_{\text{syst}}$ is the system's area, and where $\Delta \Gamma^{\text{int}}$ denotes the \emph{differential integrated rate} (DIR), defined as
\be
\Delta \Gamma^{\text{int}}=(1/2)\int_0^{\infty} \Gamma_+ (\omega) - \Gamma_- (\omega) \text{d} \omega . \label{DIR_def}
\ee
The non-linearity of the response ($\Delta \Gamma^{\text{int}}\!\sim\!E^2$) highlights its dissipative nature~\cite{Bennett}:~the depletion rate $\Gamma_{\pm} (\omega)$ reflects the power $P_{\pm}(\omega)\!=\!\hbar \omega\Gamma (\omega)$ absorbed by the system~\cite{Bennett}, which can be directly measured in (non-interacting) ultracold-atom systems through band mapping,~i.e.~by measuring the dynamical repopulation of the Bloch bands~\cite{Aidelsburger}. 

The result in Eq.~\eqref{main_result} directly results from the universal Kramers-Kronig relations, from which one can derive the more general result~\cite{Bennett,Souza,tran2017}
\begin{equation}
\Delta \Gamma^{\text{int}}/A_{\text{syst}}=(2\pi E^2/\hbar) \, \sigma_{xy} , \label{final_dichroism_rate}
\end{equation}
where $\sigma_{xy}$ denotes the transverse conductivity of the system. The result in Eq.~\eqref{main_result} then directly follows from Eq.~\eqref{final_dichroism_rate}, upon invoking the TKNN result~\cite{Thouless}, $\sigma_{xy}= -(1/h)  \nu$, relating the electric Hall conductivity of a 2D Chern insulator to its Chern number $\nu$; we set the charge $e\!=\!1$ to equally treat neutral atomic gases~\cite{Cooper_review,Dalibard_review,Goldman_review}. Care is however required when considering Chern insulating systems with boundaries~\cite{tran2017,bianco2011,Souza}:~the edge-states contribution, which is absent when considering periodic boundary conditions, is found to exactly cancel the bulk response in Eq.~\eqref{main_result}, hence leading to an overall trivial result: $\Delta \Gamma^{\text{int}}_{\text{bulk}}+\Delta \Gamma^{\text{int}}_{\text{edge}}\!=\!0$. This effect is compatible with the fact that the total conductivity necessarily vanishes in confined and isolated insulating systems. Several schemes were proposed in Ref.~\cite{tran2017} to remove the edge-state contributions, based on cold-atom technologies, hence allowing one to recover the quantization law [Eq.~\eqref{main_result}] in finite systems with boundaries.

Here, we explore the applicability of the quantization law in Eq.~\eqref{main_result} in the context of two-dimensional gases subjected to a uniform (synthetic) magnetic field, and prepared in the quantum Hall (QH) regime~\cite{Cooper_review,Dalibard_review,Goldman_review}. In this case, the spectrum of the probed system is constituted of an infinite number of Landau levels (LLs), which is in contrast to the Chern-insulator case~\cite{tran2017}, where the relevant spectrum can be restricted to a limited number of (dispersive) bands whose Chern numbers sum up to zero~\cite{Bernevig}. In the LL framework, the integration over the drive frequency $\omega$ required by the definition of the DIR in Eq.~\eqref{DIR_def} is \emph{a priori} unbounded; this questions the possibility of accessing the quantization law in Eq.~\eqref{main_result} in experiment. As we recall below, the LL structure is associated with selection rules that highly constrain the transitions generated by a circular drive [Eq.~\eqref{eq:circulardrive}]; this leads to a simple analysis of the quantized DIR that is associated with a filled LL in terms of Rabi oscillations. As a corollary, our analysis suggests that this quantization of the DIR could be directly revealed by addressing a single Rabi oscillation in the bulk of the QH system. Moreover, we recover that the edge of the system is also associated with a quantized DIR, and we show that this edge effect could be revealed by addressing a single Rabi oscillation at the boundary of the system. In both cases, the corresponding Rabi frequency is shown to be constrained by the Chern number of the LL ($\nu\!=\!1$). This result provides an alternative method by which the topological nature of the bulk can be detected by performing a local measurement at the boundary of the system.
   
 \paragraph*{Landau levels subjected to a circular perturbation}
 
We study the response of a completely filled LL in the presence of an external circular drive~[Eq.~\eqref{eq:circulardrive}]. The physical systems of interest are those describing fermions of mass $M$, which move in the $x$-$y$ plane in the presence of a constant and perpendicular magnetic field $\mathbf{B}=B\mathbf{1}_z$. The corresponding single-particle Hamiltonian gives rise to equispaced LLs, whose energies are multiples of the cyclotron energy, $E_n\!=\!\hbar\omega_c \times n$, where we introduced the cyclotron frequency $\omega_c=B/M$, the Landau level index $n= 0,1,2,\ldots$, and where we set the zero-point energy $E_0\!=\!0$; besides each of these levels has a degeneracy indexed by a non-negative integer $m=0,1,2\ldots$, which is related to the eigenvalue of the angular-momentum operator, $\langle \hat L_z \rangle\!=(n-m)\hbar$. Of particular interest here are the states in the lowest Landau level (LLL) $n\!=\!0$, given by~\cite{Yoshioka}
 \begin{align}
 \langle \mathbf{r}\vert n\!=\!0; m\rangle=C_m z^{m} e^{-\vert z \vert^2/4},
 \label{eq:lll}
 \end{align}
 where $z\!=\!(x\!-\!iy)/l_B$, $C_m$ is a normalization constant, and where $l^2_B\!=\!\hbar/B$ is the magnetic length. We recall that the mean radius of these states is given by $r_m\!=\!\sqrt{2l^2_B(m+1)}$. Similar expressions are known for the higher-energy LL states~\cite{Yoshioka}.
 
\begin{figure}
{\scalebox{0.19}
{\includegraphics{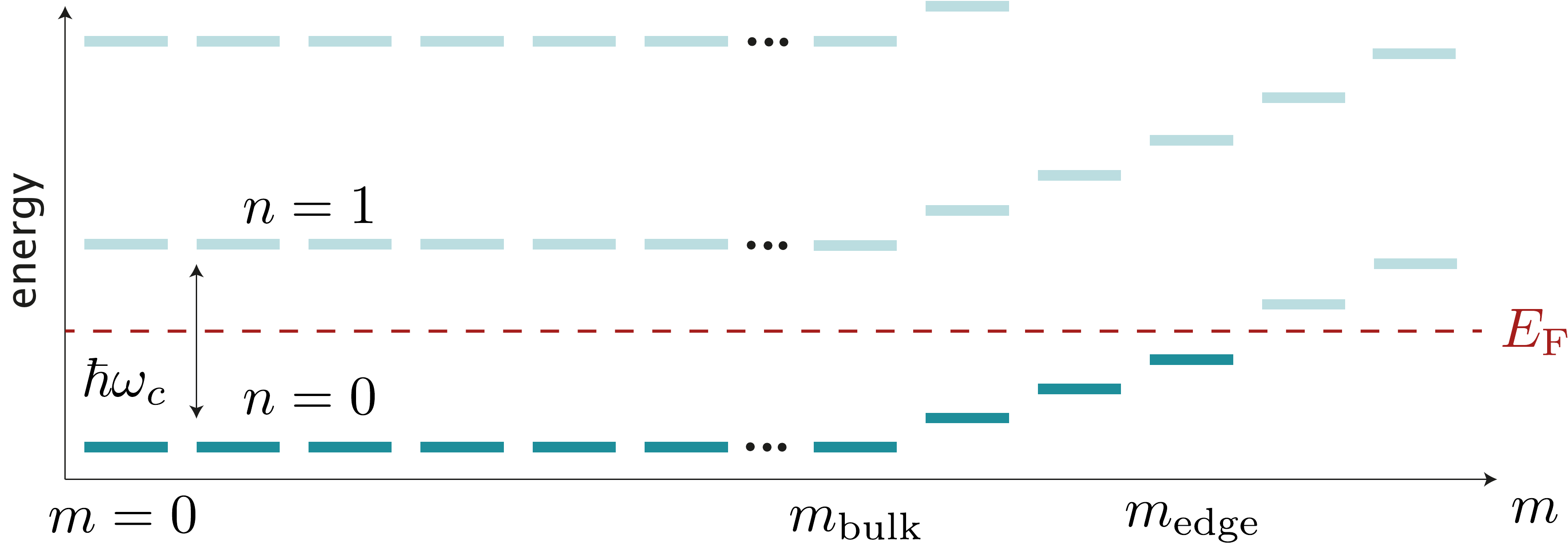}}} 
\caption{Sketch of Landau levels $(n,m)$ in a box potential.}\label{Fig_Landau_edge}
\end{figure}

 
In the following, we consider a finite system of circular geometry, and we set the Fermi energy within the first bulk energy gap~\cite{Halperin1982}; see Fig.~\ref{Fig_Landau_edge} for a sketch. In this configuration, the LLL is completely filled with non-interacting fermions, up to the angular-momentum index denoted $m_{\text{edge}}\!\gg\! 1$; the mean area of the system is thus given by $A_\text{syst}\!=\!\pi r_{m_{\text{edge}}}^2$, where $r_{m_{\text{edge}}}\!=\!\sqrt{2l^2_B(m_{\text{edge}}+1)}$ denotes the mean radius of the last occupied state. We will first assume that the system is confined in a box potential~\cite{Corman2014}, which allows for a clear distinction between bulk and edge states: the bulk states will correspond to those populated LLL states that are not significantly affected by the confining potential. We introduce the notation $m_{\text{bulk}}$ to designate the number of bulk states; the number of populated edge states is then $N_{\text{edge}}\!=\!m_{\text{edge}}-m_{\text{bulk}}$ in these notations [Fig.~\ref{Fig_Landau_edge}]. We stress that our results are valid for a wide class of potentials, including harmonic traps, as discussed later.

In this filled-LLL configuration, the system exhibits the QH effect~\cite{Yoshioka}:~the bulk Hall conductivity is quantized as $\sigma_{xy}\!=\!-(1/h)$. According to the bulk-edge correspondence~\cite{Bernevig}, this result reflects the fact that a single edge mode is populated and that the Chern number associated with the LLL is $\nu\!=\!1$. Following Ref.~\cite{tran2017}, we instead study the excitation rate of this QH system in response to the circular  drive $\hat V_{\pm}(t)$ in Eq.~\eqref{eq:circulardrive}. Importantly, the latter perturbation only couples a limited set of LLL states, as dictated by the following selection rules:
\begin{align}
&\langle \tilde n ; \tilde m \vert \hat x + i \hat y \vert n=0 ; m \rangle = \delta_{\tilde n,0} \delta_{\tilde m,m+1} \times l_{B}\sqrt{2 (m+1) }, \notag \\
&\langle \tilde n ; \tilde m \vert \hat x - i \hat y \vert n=0 ; m \rangle = \delta_{\tilde n,0} \delta_{\tilde m,m-1} \times l_{B}\sqrt{2 m } \notag \\
& \qquad \qquad \qquad \qquad \qquad \quad -\delta_{\tilde n,1} \delta_{\tilde m,m} \times i l_{B}\sqrt{2} .\label{selection_rules}
\end{align}
While the ``positive" orientation $\hat V_+(t)$ can only generate transitions within the LLL, the opposite orientation $\hat V_-(t)$ can also trigger inter-level excitations. 

Computing the DIR in Eq.~\eqref{DIR_def} for the total system leads to a vanishing net result, consistent with the vanishing Hall conductivity for any bounded system; this result can readily be obtained for the LLL subjected to a harmonic confinement [see Appendix]. We therefore now turn to the computation of bulk and edge responses,  $\Delta \Gamma^{\text{int}}_{\text{bulk/edge}}$, based on the selection rules in Eq.~\eqref{selection_rules}. We stress that the latter are only strictly valid for ideal (unconfined) LL states, so our results apply assuming that the LL states are only slightly perturbed by the confinement. For a harmonic confinement potential, $V(r) = \frac{1}{2}M \omega_{\rm T}^2|\mathbf{r}|^2$, the relevant energy eigenstates are the Fock-Darwin states [see Appendix], which take the form of Eq.~\eqref{eq:lll} but with the magnetic length rescaled to $\tilde{l}_B^2 = \hbar/(M\sqrt{ \omega_c^2 + 4 \omega_{\rm T}^2})$. From this, one sees that the selection rules hold accurately provided  $\omega_T\!\ll\!\omega_c$. 
More generally, for any potential $V(r)$, which is smooth on the length-scale of the magnetic length such that it is accurate to represent it locally by a Taylor expansion, the condition is that the local curvature satisfies
$V''(r) \ll \hbar\omega_c/l_B^2$. 

\paragraph*{Quantization of the DIR in the bulk}

\begin{figure}
{\scalebox{0.19}
{\includegraphics{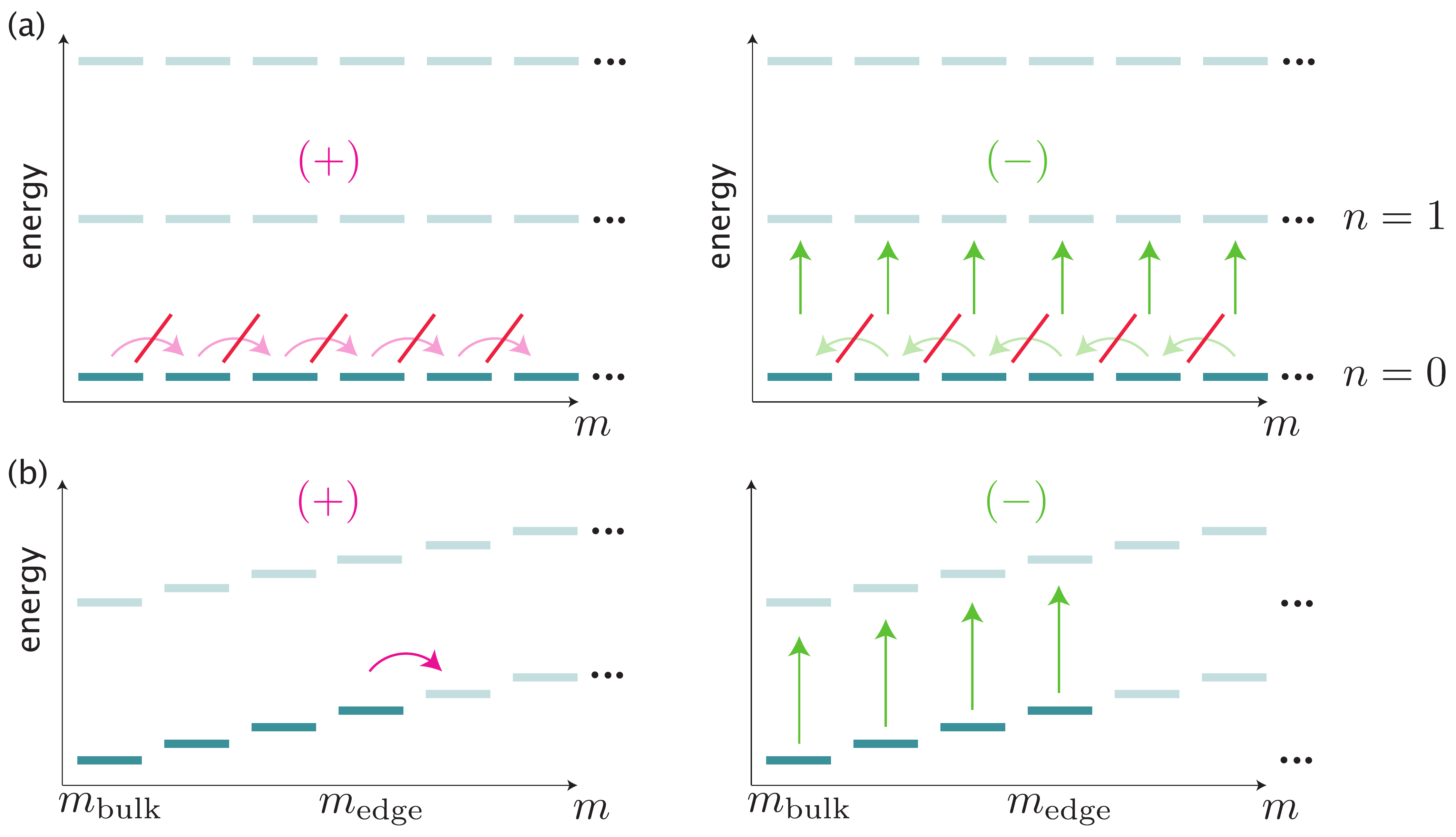}}} 
\caption{Possible transitions induced by the drive $\hat V_{\pm}(t)$ in Eq.~\eqref{eq:circulardrive}, and involving the initially occupied bulk states (a) and edge states at zero temperature (b), according to Eq.~\eqref{selection_rules} and Pauli exclusion principle (red hatching). The $(\pm)$ symbols refer to the chirality of the drive.}\label{Fig_Landau_selection_bulk}
\end{figure}

We now calculate the DIR defined in Eq.~\eqref{DIR_def}, and first focus on the contribution of the bulk states as defined above [i.e.~$m\!\le m_{\text{bulk}}$, see Fig.~\ref{Fig_Landau_edge}]. We notice that the selection rules [Eq.~\eqref{selection_rules}], combined with Pauli exclusion's principle, highly constrain the possible transitions involving the initially occupied bulk states. In particular, we find that only $\hat V_-(t)$ can activate excitations, and that the latter strictly correspond to $(n\!=\!0,m)\!\rightarrow\!(n\!=\!1,m)$ processes [see Fig.~\ref{Fig_Landau_selection_bulk}(a)]; this sets $\Gamma_+(\omega)\!=\!0$ in this case. We disregard higher-order excitations, involving LL states with $n>1$, which is a valid assumption for observation times much smaller than the Rabi period (to be defined below). Consequently, the total excitation rate in response to the drive $\hat V_- (t)$ is obtained by analyzing ``$m_{\text{bulk}}$" independent copies of the following 2-level system
\begin{equation}
\hat H(t) = \hbar \omega_c \vert 1 \rangle \langle 1 \vert + \left ( \hbar \Omega_{-} e^{-i \omega t} \vert 1 \rangle \langle 0 \vert + \text{h.c.} \right ), 
\end{equation}
where $ \vert 0,1 \rangle\!\equiv\!\vert n\!=\!0,1; m\rangle$ denote two coupled Landau levels, and where the Rabi frequency $\Omega_{-}\!=\!  (E l_{B}/\hbar) \times (i\sqrt{2})$ is deduced from Eq.~\eqref{selection_rules}.  The total excitation rate $\Gamma_-(t)$ is then given by the Rabi formula~\cite{Cohen_book}
\begin{equation}
\Gamma_-(\omega)\approx t \vert \Omega_- \vert^2 \text{sinc}^2 \left [t(\omega - \omega_c)/2 \right ] \times m_{\text{bulk}}, \notag
\end{equation}
where we assumed $t \ll 1/\vert \Omega_- \vert$. One can now estimate the DIR in Eq.~\eqref{DIR_def} by integrating $\Gamma_-(\omega)$ over the frequency $\omega$, 
\begin{align}
\Delta \Gamma^{\text{int}}_{\text{bulk}} &= - \frac{1}{2}\int_0^{\infty}\Gamma_-(\omega) \text{d} \omega \approx - \pi \, \vert \Omega_- \vert^2 \times m_{\text{bulk}}  \label{bulk_DIR} \\
&= - 2 \pi (E/\hbar)^2 l_{B}^2 \times m_{\text{bulk}}. 
\end{align}
Noting that $A_{\text{syst}}\!=\!2\pi l_{B}^2 (m_{\text{edge}}+1)$ finally leads to the quantized result
\begin{equation}
\Delta \Gamma^{\text{int}}_{\text{bulk}}/A_{\text{syst}} = (E/\hbar)^2 \times (-1) , \label{bulk_quantized}
\end{equation}
upon taking the thermodynamic limit $m_{\text{bulk}}\!\sim\!m_{\text{edge}}\!\rightarrow\!\infty $. We point out that this result is compatible with the laws presented in Eqs.~\eqref{main_result}-\eqref{final_dichroism_rate} and the quantized Hall conductivity associated with the LLL, $\sigma_{xy}\!=\!-(1/h)$, as established by the topological Chern number $\nu\!=\!1$ characterizing the LLL. We note that this quantized bulk response is robust at non-zero temperature, as long as the latter remains small compared to $\hbar \omega_c / k_B$.

This analysis suggests a simple protocol to reveal the quantization of $\Delta \Gamma^{\text{int}}_{\text{bulk}}$ in Eq.~\eqref{bulk_quantized}. Since the DIR in Eq.~\eqref{bulk_DIR} only depends on the Rabi frequency $\vert \Omega_- \vert$ and the index $m_{\text{bulk}}$, one could readily access this quantity by driving a single Rabi oscillation. Indeed, when considering a resonant drive $\hat V_{-} (t)$ with $\omega\!=\!\omega_c$, the number of excited particles would read~\cite{Cohen_book}
\begin{equation}
N_-(t) \approx m_{\text{bulk}}\left ( \vert \Omega_- \vert t \right )^2, \quad t \ll 1/\vert \Omega_- \vert, \label{eq:Nplus}
\end{equation}
hence giving directly access to the Rabi frequency $\vert \Omega_-\vert$ (and $m_{\text{bulk}}$), and incidentally, to the quantized $\Delta \Gamma^{\text{int}}_{\text{bulk}}$ via Eq.~\eqref{bulk_DIR}. Using Eqs.~\eqref{bulk_DIR}-\eqref{eq:Nplus}, one can introduce the following ``quantization indicator" for this protocol
\begin{align}
C=\dfrac{\pi N_-(t)}{A_\text{syst} (t E/\hbar)^2}\approx\dfrac{- \Delta\Gamma^\text{int}_\text{bulk}}{A_\text{syst}(E/\hbar)^2}\rightarrow+1 ,
\label{eq:quantbulk}
\end{align}
which highlights the manner by which the number of excitations $N_-(t)$ is constrained by the quantization law~\eqref{bulk_quantized}, and hence, by the topology of the populated LLL.

\begin{figure}
{\scalebox{0.2}
{\includegraphics{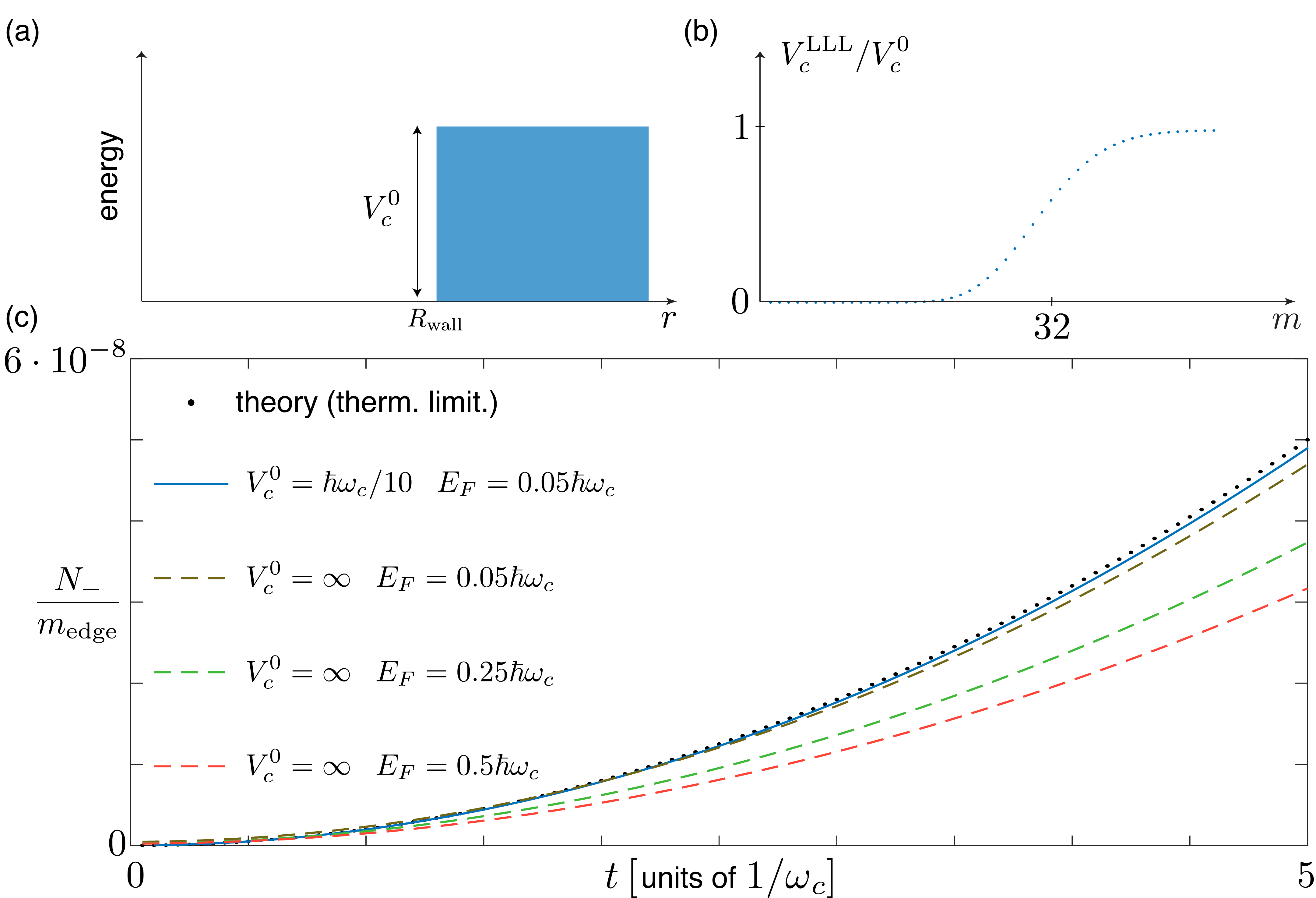}}} 
\caption{(a) Sketch of the radial box confinement, with finite depth $V_c^0$ and radius $R_\text{wall}$. (b) The effective potential  $V_c^\text{LLL}$, i.e.~the first-order correction to the energies in the LLL, for a trap radius $R_\text{wall}\!=\!8l_B$ and a finite depth $V_c^0\!\ll\!\hbar \omega_c$. (c) Excited fraction $N_-(t)/m_{\text{edge}}$, i.e.~number of excited particles normalized by the total number of particles, upon subjecting the system to the (resonant) circular drive $\hat V_{-} (t)$, for different depth $V_c^0$ and Fermi energies $E_F$. The infinite depth case, $V_c^0\!=\!\infty$, corresponds to setting the radius of the simulation box to be equal to $R_\text{wall}$. The quantization indicator in Eq.~\eqref{eq:quantbulk}, as extracted from the top to bottom curves, are $(+1)$, $(+0.97)$, $(+0.93)$, $(+0.74)$ and $(+0.63)$, respectively. All computations were done with a driving strength $E\!=\!3.2\times 10^{-5}\hbar \omega_c/l_B$; the system sizes used are $R_\text{wall}=6.3l_B$ for $V_c^0=\hbar \omega_c/10$, and $R_\text{wall}=9.5l_B$ for $V_c^0=\infty$; the different radii $R_\text{wall}$ were chosen in order to have the same number of bulk states ($m_\text{bulk}$).}\label{Nplus}
\end{figure}

We now numerically explore the viability of this simple protocol, by analyzing the excitation fraction $N_-(t)/m_{\text{edge}}$ for different trap strength and initial fillings. Specifically, we consider the LL states for a circular box potential of depth $V_c^0$ [Fig.~\ref{Nplus}(a)], and time-evolve the system in the presence of the (resonant) circular drive $\hat V_{-} (t)$; the initial filling of the system is set by the Fermi energy $E_F\!>\!0$, which also sets the difference $m_{\text{edge}}-m_{\text{bulk}}$ according to the effective potential~\cite{Macaluso2017} shown in Fig.~\ref{Nplus}(b). First, we validate the theoretical prediction in Eq.~\eqref{eq:Nplus}, by comparing it to the numerics obtained for the quasi-ideal case where the confinement depth $V_c^0\!\ll\!\hbar \omega_c$; see the dotted and full lines in Fig.~\ref{Nplus}(c), which are found to correspond to the quantization indicator $C\!=\!1$ and $C\!=\!0.97$, respectively; note that the dotted line was obtained by setting $m_{\text{bulk}}\!=\!m_{\text{edge}}$ in Eq.~\eqref{eq:Nplus}, which is only strictly valid in the thermodynamic limit. Then, we analyze how this behavior is modified as one further perturbs the LL states. To do so, we consider an infinitely deep box $V_c^0\!=\!\infty$, and calculate the excitation fraction for increasing values of the Fermi energy. One should note that increasing the latter, for a given system size, also leads to a deviation from the thermodynamic-limit regime (where $m_{\text{bulk}}\!\approx\!m_{\text{edge}}$). The resulting excitation fractions, shown by dashed lines in Fig.~\ref{Nplus}(c), correspond to the quantization indicators $C\!=\!0.93$, $C\!=\!0.74$ and $C\!=\!0.63$, respectively. This numerical analysis confirms the validity of this protocol, whenever the populated states are only slightly perturbed by the trapping potential ($E_F\!\approx\!E_0$) and the thermodynamic-limit ($m_{\text{bulk}}\!\sim\!m_{\text{edge}}\!\rightarrow\!\infty $) is reached.

\paragraph*{Quantization of the DIR on the boundary}

A complementary analysis can be performed on the boundary of the system.  Here, in contrast to the bulk-contribution study, nonzero temperature has an important impact on the results. Let us first consider the case of zero temperature, where Pauli exclusion's principle and the selection rules in Eq.~\eqref{selection_rules} lead to a very limited set of possible transitions involving the initially occupied edge states [Fig.~\ref{Fig_Landau_selection_bulk}(b)]. In this case, an intra-LLL transition, which can be addressed by driving the system with the drive $\hat V_+(t)$, is now possible in the vicinity of the edge. Since the related Rabi frequency $\Omega_+\!=\!(E l_{B}/\hbar)\sqrt{2 (m_{\text{edge}}+1)}$ scales with the system size [see Eq.~\eqref{selection_rules}], and since the system typically only displays a limited set of edge states, $N_{\text{edge}}\!\ll\!m_{\text{edge}}$, one can safely restrict our analysis to this specific orientation of the drive (i.e.~$\Gamma_+\gg\Gamma_-$). Applying the Rabi formula as above, we find that the DIR now satisfies
\begin{align}
\Delta \Gamma^{\text{int}}_{\text{edge}} &\approx (1/2)\int_0^{\infty}\Gamma_+(\omega) \text{d} \omega\approx \pi \,  \vert \Omega_+ \vert^2 \label{edge_DIR}\\
&=2 \pi (E/\hbar)^2 l_{B}^2 (m_{\text{edge}}+1),    \notag
\end{align}
which leads to the edge-counterpart of Eq.~\eqref{bulk_quantized},
\begin{equation}
\Delta \Gamma^{\text{int}}_{\text{edge}}/A_{\text{syst}} = (E/\hbar)^2 \times (+1) , \label{edge_quantized}
\end{equation}
when considering the thermodynamic limit $m_{\text{edge}} \rightarrow \infty$. We recover a quantization law associated with an edge response, and the fact that the edge and bulk contributions exactly cancel each other, i.e.~ $\Delta \Gamma^{\text{int}}_{\text{edge}}\!=\! - \Delta \Gamma^{\text{int}}_{\text{bulk}}$. 

We note that the quantization law in Eq.~\eqref{edge_quantized} could be directly probed using a minimal setting:~Driving a single Rabi oscillation at the boundary of the system, with a drive $\hat V_+(t)$ whose frequency $\omega$ is set on resonance with respect to the relevant transition $(n\!=\!0,m_{\text{edge}})\!\rightarrow\!(n\!=\!0,m_{\text{edge}}+1)$, could be exploited to extract the Rabi frequency $\vert \Omega_+ \vert $ entering Eq.~\eqref{edge_DIR}, hence offering a direct and local measure of the quantized $\Delta \Gamma^{\text{int}}_{\text{edge}}$ in Eq.~\eqref{edge_quantized}. In this case, the corresponding drive frequency will typically be very small, as can be estimated by calculating the effective potential felt by the LLL in the presence of confinement~\cite{Macaluso2017}; see Fig.~\ref{Nplus}(b). \\


The latter edge-contribution study relies on supposing that the system is at zero temperature. In the presence of nonzero temperature (but small compared to $\hbar \omega_c / k_B$), the edge region has a smooth fall-off of the Fermi distribution over many $m$ orbitals, which results in having both $\Gamma^{\text{int}}_{+}\!\sim\!\Gamma^{\text{int}}_{-}$ large. Importantly, the quantization law still exists for the DIR, as can be obtained by repeating the calculations above now for partially filled states in the vicinity of the Fermi energy. Indeed, writing $M$ the number of states with fractional thermal occupation, we find the DIR
\begin{align}
\Delta \Gamma^{\text{int}}_{\text{edge}} &= (1/2)\int_0^{\infty}\Gamma_+(\omega)-\Gamma_-(\omega) \text{d} \omega \\
&\approx 2 \pi (E/\hbar)^2 l_{B}^2 (m_{\text{edge}}-M/2 +1),    \notag
\end{align}
which now takes the response of both drive orientations $\hat V_{\pm}$ into account. This circular dichroism yields the quantization
\begin{equation}
\Delta \Gamma^{\text{int}}_{\text{edge}}/A_{\text{syst}} = (E/\hbar)^2 \times (+1) , \qquad m_{\text{edge}} \rightarrow \infty ,\label{edge_quantized_2}
\end{equation}
in the thermodynamic limit. This demonstrates the robustness of the quantized edge response in the presence of temperature, through a genuine circular-dichroism effect.

Crucially, the spatial localization of LLs around their radius $r_m\!=\!\sqrt{2l_B^2(m+1)}$, allows for a separation between bulk and edge states, and hence, for the possibility of individually (and locally) probing bulk and edge contributions, $\Delta \Gamma^{\text{int}}_{\text{bulk/edge}}$. This is in contrast with Chern insulators~\cite{tran2017}, where bulk states are delocalized over the sample, and thus significantly overlap with the edge states. We note that a bulk-edge separation could still be found in the Chern-insulator case, when trapping the system in a sufficiently weak and smooth trap (such that the edge modes are spatially separated from the bulk~\cite{Goldman_ETH,Buchhold,Goldman_edge_bis}).

\paragraph*{Discussion}

These results could be investigated using ultracold Fermi gases subjected to uniform (synthetic) magnetic fields, as can be realized through rotation~\cite{Cooper_review} or by applying laser-induced gauge fields~\cite{Dalibard_review,Goldman_review}. The effective cyclotron frequency is typically of the order of $\omega_c\!\sim\! 100 \text{Hz} $, however larger values (up to $\omega_c\!\sim\! 100 \text{kHz} $) could be achieved~\cite{Cooper_review,Dalibard_review,Goldman_review}. To achieve the LLL limit requires both the chemical potential and the temperature to be less than $\hbar\omega_c$~\cite{Cooper_review}. This is challenging for small $\omega_c$, but has been achieved in experiments on rotating bosons~\cite{schweikhard} even with $\omega_c \simeq 2\pi \times 16\mbox{Hz}$. For a (one-component) Fermi gas, the condition on chemical potential restricts the 2D particle density of the filled LLL to $\rho = \omega_c M/h$, potentially requiring the preparation of a gas at very small densities. The circular drive in Eq.~\eqref{eq:circulardrive} could be generated by shaking the gas circularly~\cite{Jotzu,Hamburg}, or by subjecting the system to time-modulated (magnetic) gradients~\cite{Xu_Ueda}. The results above find their origin in the selection rules [Eq.~\eqref{selection_rules}], which remain valid as long as the LL states are only slightly affected
by the confinement. For smooth potentials this corresponds to sufficiently small
curvature of the potential, while for a hard-wall potential the condition is that the relevant states are far enough from the hard wall that their energy shifts are small compared to $\hbar \omega_c$.
 
These results could be applied to other settings, e.g.~graphene-type systems exhibiting ``relativistic" LLs, where similar absorption properties and selection rules are found~\cite{Gullans_2017,Ghazaryan_2017}. While our results apply to any non-interacting Fermi gas subjected to a (synthetic) uniform magnetic field, we point out that their implication to solid-state experiments~\cite{footnote_shake} is not immediate, as the effects of disorder, non-circular boundaries and interactions are likely to alter the Rabi oscillations, and hence, the quantized properties revealed in the work.\\
 
Note added in proof: An experimental observation of quantized circular dichroism was recently reported in Ref.~\cite{Asteria2018}. \\
 

\paragraph*{Acknowledgments} We wish to thank J.~Dalibard, whose curiosity triggered the present research. Discussions with M.~Aidelsburger, I.~Carusotto, A.~Dauphin, A.~G.~Grushin, G.~Jotzu, H.~M.~Price, C.~Repellin, C.~Weitenberg, and P.~Zoller are also acknowledged. Work in Brussels in supported by the FRS-FNRS (Belgium) and the ERC Starting Grant TopoCold. NRC acknowledges support of EPSRC Grants EP/K030094/1 and EP/P009565/1.


\section*{Appendix: Harmonic Confinement}

\label{appendix}

The features described in the text can  be worked out in  detailed for the special case of harmonic confinement potential. 

\paragraph{Single particle wavefunctions}
The Hamiltonian for a single particle takes the form
\begin{eqnarray}
\hat{H} & = & \frac{1}{2M}\left[\hat{\mathbf{p}}-\mathbf{A}(\hat{\mathbf{r}})\right]^2 + \frac{1}{2} M \omega_T^2 |\hat{\mathbf{r}}|^2 ,
\end{eqnarray}
where we set the particle charge to be one. Taking the symmetric gauge, $\mathbf{A}(\mathbf{r}) = \frac{1}{2}B(-y,x)$, the Hamiltonian can be written
\begin{eqnarray}
\label{eq:ham}
\hat{H} & = & \frac{1}{2M}\hat{\mathbf{p}}^2 + \frac{1}{2} M \left(\omega_T^2+ \frac{1}{4}\omega_c^2\right) |\hat{\mathbf{r}}|^2 - \frac{1}{2}\omega_c \hat{L}_z ,
\end{eqnarray}
where $\omega_c \equiv B/M$ is the cyclotron frequency and $\hat{L}_z \equiv \hat{\mathbf{r}}\times\hat{\mathbf{p}}\cdot \mathbf{z}$ is the angular momentum. The angular momentum is a constant of the motion, so can be taken as $L_z = \hbar m_\ell$, with $m_\ell$ integer. The energy eigenstates are those of a circular oscillator
with natural frequency
${\tilde \omega}_T \equiv \sqrt{\omega_T ^2 + \omega_c^2/4}$ and oscillator length ${\tilde a} \equiv \sqrt{\hbar/M{\tilde \omega}_T}$. Working in polar coordinates $\mathbf{r} \to (r,\phi)$, and denoting the radial quantum number $n_r$ with $m_\ell$ the angular momentum, these energy eigenstates are
\begin{equation}
\psi_{n_r,m}(r,\phi) \propto r^{|m_\ell|} e^{im_\ell\phi} e^{-r^2/2{\tilde a}^2}  {}_1F_1(-n_r,|m_\ell|+1;r^2/{\tilde a}^2),
\label{eq:fdstates}
\end{equation}
where $_1F_1(a,b;c)$ is the Kummer hypergeomeric function~\cite{fluggepracticalqm}.
The corresponding eigenvalues of (\ref{eq:ham}) are
\begin{equation}
E_{n_r,m_\ell} = \hbar{\tilde \omega}_T (n_r + |m_\ell | + 1) - \frac{1}{2}\hbar\omega_c m_\ell \,.
\end{equation}
These are the Fock-Darwin states, written in cylindrical polar co-ordinates,  and the associated energy spectrum.

The states in the lowest Landau level correspond to those with $n_r=0$ and $m\geq 0$. For these the wavefunctions (\ref{eq:fdstates}) are simply
\begin{eqnarray}
\label{eq:fdstateslll}
\psi_{n_r,m_\ell}(r,\phi) &  \propto & r^{m_\ell} e^{im_\ell\phi} e^{-r^2/2{\tilde a}^2}  \\
& \propto & {\tilde z}^{m_\ell} e^{-|{\tilde z}|^2/4} ,
\end{eqnarray}
with ${\tilde z} \equiv (x+iy)/{\tilde l}_B$ and 
\begin{equation}
{\tilde l}_B^2 \equiv \frac{1}{2} {\tilde a}^2  = \frac{\hbar}{2M{\tilde \omega}_T}= \frac{\hbar}{M\sqrt{4\omega_T^2 + \omega_c^2}}\,.
\end{equation}
 Thus, for $\omega_T \ll \omega_c$, the lengthscale ${\tilde l}_B$ tends to the magnetic length $l_B = \sqrt{\hbar/M\omega_c}$ and the selection rules [Eq.~(6)] of the text are satisfied.
  Similarly, the energies of states in the lowest Landau level, $n_r=0$ and $m_\ell\geq 0$, are
\begin{equation}
E_{n_r=0,m_\ell} = \hbar{\tilde \omega}_T  + \frac{1}{2}\hbar(2{\tilde \omega}_T -\omega_c) m_\ell\,.
\end{equation}
For $\omega_T \ll \omega_c$ these tend to 
\begin{equation}
E_{n_r=0,m_\ell} \to \frac{1}{2}\hbar\omega_c   +\frac{\hbar \omega_T^2}{\omega_c} (m_\ell+1) = \frac{1}{2}\hbar\omega_c + \frac{1}{2} M \omega_T^2 r_{m_\ell}^2 ,
\end{equation}
with $r_{m_\ell} = \sqrt{2 l_B^2(m_\ell+1)}$ the mean radius of the state.

\paragraph{Differential Integrated Response}

Consider now an $N$-particle system with harmonic confinement and interparticle interactions
\begin{eqnarray}
\nonumber
\hat{H} & = & \sum_{i=1}^N \left\{\frac{1}{2M}\left[\hat{\mathbf{p}}_i-\mathbf{A}(\hat{\mathbf{r}}_i)\right]^2 + \frac{1}{2} M \omega_T^2 |\hat{\mathbf{r}}_i|^2 \right\}
\\
  & & + \frac{1}{2}\sum_{i\neq j} V(\hat{\mathbf{r}}_i- \hat{\mathbf{r}}_j)\,.
  \end{eqnarray}
The harmonic nature of the confinement, and the fact that the interactions depend only on the interparticle separation, allow one to separate the center-of-mass position and momentum
\begin{eqnarray}
\hat{\mathbf{R}} & = & \frac{1}{N} \sum_{i=1}^N \hat{\mathbf{r}}_i \qquad \hat{\mathbf{P}} =  \sum_{i=1}^N \hat{\mathbf{p}}_i ,
\end{eqnarray}
which behave as canonically conjugate pair. 
The center-of-mass Hamiltonian 
\begin{equation}
\hat{H}_{\rm CoM} = \frac{1}{2M_N}\left[\hat{\mathbf{P}}-Q\mathbf{A}(\hat{\mathbf{R}})\right]^2 + \frac{1}{2} M \omega_T^2 |\hat{\mathbf{R}}|^2 ,
\end{equation}
is that of a particle of mass $M_N \equiv NM $ and charge $Q=N$ moving in the harmonic well.

The  long-wavelength driving,  required to calculate the DIR, couples only to the center-of-mass coordinates, via
\begin{equation}
\Delta \hat{H} = - Q \mathbf{E}\cdot  \hat{\mathbf{R}} \,.
\end{equation}
Thus, the DIR for the interacting many-particle system can be found from the analysis of a single-particle problem associated with the center-of-mass.
The current response is readily obtained by considering the Heisenberg equations of motion of $\hat{\mathbf{R}}$ and $\hat{\mathbf{P}}$. These follow the classical equations of motion
\begin{eqnarray}
\ddot{X}
 & = & \omega_c \dot{Y} - \omega_T^2 X +\frac{Q}{M_N} 
E_x - \eta \dot{X}, \\
\ddot{Y}
 & = & -\omega_c \dot{X} - \omega_T^2 Y +\frac{Q}{M_N} 
E_y - \eta \dot{Y} ,
\end{eqnarray}
where a damping term $\eta> 0$ has been introduced to keep track of causality.
Writing 
\begin{eqnarray}
E_x & = & \Re\left[ E_+ e^{i\omega t} + E_- e^{-i\omega t} \right], \\
E_y & = & \Re\left[ -i E_+ e^{i\omega t} + i E_- e^{-i\omega t} \right] ,
\end{eqnarray}
in terms of the right/left circular polarization amplitudes $E_\pm$,  one finds that the complex amplitudes $X_\pm \equiv X \pm i Y$ are
\begin{equation}
X_\pm \equiv \chi_\pm E_\pm = \frac{(Q/M_N) 2 E_\pm }{\omega_T^2 \mp \omega\omega_c - \omega^2 + i \eta \omega}.
\end{equation}
The net power absorbed is
\begin{eqnarray}
P  & = & \frac{Q}{2} \Re \left[E_+^* \dot{X}_+ + 
E_-^* \dot{X}_-\right] ,\\
 & = &  \frac{Q}{2}\omega \left[|E_+|^2 \Im(-\chi_+)
+ |E_-|^2 \Im(-\chi_-)\right] .
\label{eq:power}
\end{eqnarray}

(i) For a static drive,
\begin{equation}
\chi_\pm(\omega = 0) = \frac{2 Q}{M_N \omega_T^2},
\end{equation}
one recovers the expected static displacement of the center-of-mass, $\mathbf{R} = \frac{Q}{M_N \omega_T^2} \mathbf{E}$. The current vanishes, so $\sigma_{xx}=\sigma_{xy}=0$ as expected for this confined system.

(ii) Consider now the power absorbed (\ref{eq:power}) at nonzero frequencies, for which 
\begin{eqnarray}
\Im ( - \chi_\pm)   & = &  \frac{(Q/M_N) \eta\omega}{(\omega_T^2\mp \omega\omega_c-\omega^2)^2 + \eta^2\omega^2}\,.
\end{eqnarray}
Defining the {\it rates} of photon absorption as $\Gamma_\pm \equiv P_\pm/(\hbar\omega)$ for the two circular polarizations, one finds
the DIR
\begin{widetext}
\begin{eqnarray}
\Delta \Gamma^{\rm int} & = & \frac{1}{2\hbar}\frac{Q^2}{M_N} |\mathbf{E}|^2  \int_0^\infty d\omega 
\left[\frac{\eta \omega}{(\omega_T^2-\omega\omega_c-\omega^2)^2 + \eta^2\omega^2}
-\frac{\eta \omega}{(\omega_T^2 + \omega\omega_c-\omega^2)^2 + \eta^2\omega^2}\right] . \label{eq:integral}
\end{eqnarray}
\end{widetext}The two contributions in (\ref{eq:integral}) can be combined as an integral along the full real axis of $\omega$
\begin{equation}
 \int_{-\infty}^\infty d\omega 
\frac{\eta \omega}{(\omega_T^2-\omega\omega_c-\omega^2)^2 + \eta^2\omega^2} \,.
\end{equation}
Computing this integral by the residue theorem shows that it vanishes for $\eta \to 0+$, leading to $\Delta \Gamma^{\rm int} \!=\!0
$ consistent with the vanishing static $\sigma_{xy}$.

 \end{document}